# A local maximum in the superconducting transition temperature of Nb-doped strontium titanate under uniaxial compressive stress

Chloe Herrera and Ilya Sochnikov

Physics Department, University of Connecticut, Storrs, Connecticut, USA 06269

**Abstract**

Superconducting strontium titanate (STO) is in the spotlight as a low carrier concentration semiconductor in proximity to polar order. Its superconducting pairing mechanism poses an open fundamental challenge, which may be resolved by controlling how strongly the superconducting and polar orders interact. Here, we show for the first time that non-monotonic superconducting transition temperature behavior in *uniaxially compressed* STO can occur, revealing a reduction followed by a peak in the superconducting transition temperature. The corresponding displacements in the lattice are extremely small, in the sub picometer range, indicating that this material approaches a singularity in electron-phonon coupling, consistent with STO transforming to the polar (ferroelectric) phase that influences the superconductivity.

**Main text**

The mechanism of superconducting pairing in doped strontium titanate (STO) is a major fundamental open question that has stimulated intense debate *(1–12)*. Several rather new experimental works have demonstrated the enhancement of superconductivity in STO via changes in the crystal composition *(13–15)* and uniaxial or epitaxial strain techniques *(16, 17)*, and have revealed the potential importance of the ferroelectric quantum phase transition to this material's mysterious electron pairing *(3, 7, 18–24, 13, 14, 10, 25, 26)*.

Herrera *et al. (16, 27)* demonstrated that *tension* in the [001] tetragonal direction in STO can lead to a large enhancement of the critical temperature. However, only some lower-doping STO samples measured in the past *(28)* showed minor enhancement on *compression*, while higher-doping samples displayed a reduction in the critical temperature *(16, 28)*. Several questions remain. Why do the higher-doping samples show only a reduction of the critical temperature under compression, and not an enhancement? Is there a potential increase in the critical temperature upon further compression of the higher-doping crystals? Is the critical temperature suppressed again once the material is in the polar (ferroelectric) phase?

In this brief letter, we provide answers to these questions and report that the critical temperature indeed enhances in higher-doping STO samples, but only under quite large compression, and that the enhancement is milder than in previous tensile-stress experiments *(16)*.

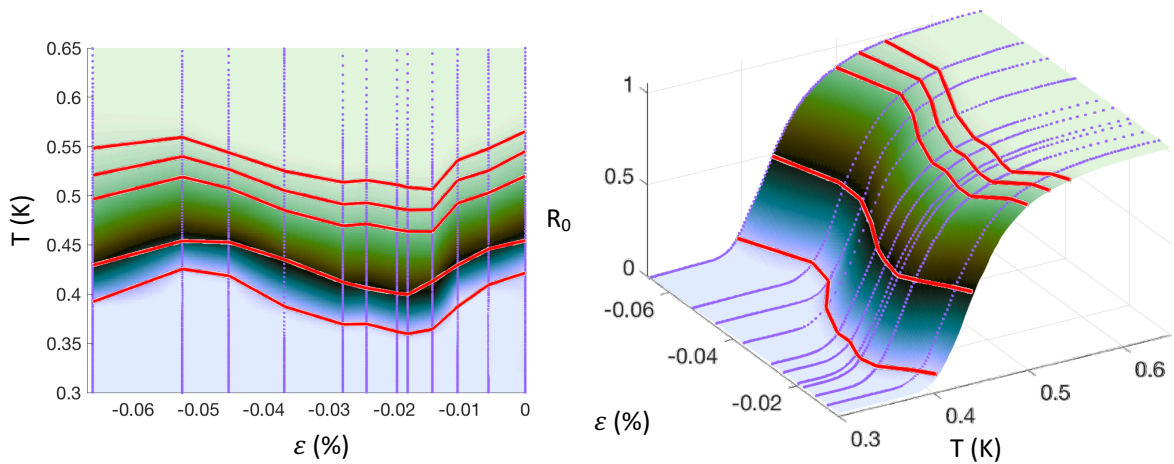

Figure 1. Superconducting transition temperature in a $Sr_1Ti_{1-x}Nb_xO_3$, x=0.004 single crystal under $[001]_c$ compressive stress. The measured compressive strain values induced by applying uniaxial stress (not measured) along this axis are shown. The red lines are the critical temperature contours defined at 10, 50, 90, 95, and 98% of the normalized resistance. The resistance was normalized to its value at 1 K, which did not change importantly under stress. For small compressive strain values, the transition temperature decreases. At values around -0.05%, a local maximum appears, followed by another downturn. The local maximum is interpreted as a passage of the sample through the polar (ferroelectric) phase transition, which moderately enhances the superconducting pairing. Beyond the polar (ferroelectric) structural transition the temperature is suppressed again.

We have performed uniaxial stress experiments in a custom dilution refrigerator with a millikelvin-compatible home-built copper strain-stress cell, all described elsewhere *(16, 27)*. The strain values along the stress direction were monitored directly on the sample using an attached miniature strain gauge. We detected the superconducting transition temperature using ultra-low noise methods, with small excitation currents of 10 μA *(27)*. Typical transition temperature data from 2 x 0.2 x 10 mm³ $Sr_1Ti_{1-x}Nb_xO_3$, x=0.004 crystals *(16)* are shown in Figure 1. The samples' carrier concentration at 300 K and mobility were $6.4 \cdot 10^{19} cm^{-3}$ and $8.32\ cm^2/Vs$.

Starting from nominally zero measured $[001]_c$ strain *(16)*, upon compression, the critical temperature begins to decrease (Figure 1, red constant-resistance contours). At strains of approximately -0.02% measured along the [001] cubic direction, a local minimum occurs (Figure 1).

Further compression reverses the trend: the critical temperature starts increasing (Figure 1). It reaches a local maximum at approximately -0.05% (Figure 1), which rather accurately matches the anticipated strains at which the ferroelectric transition in undoped samples was observed *(29)*. Even further increase in strain leads to a reduction of the critical temperature (Figure 1), in agreement with transitioning to the polar phase which has been shown to result in the hardening of the relevant ferroelectric phonon modes *(29)*. This inverse correlation between the critical temperature and the ferroelectric phonon energy is believed to occur in certain theoretical models proposing that these are the most relevant phonons for the electron-electron pairing *(3, 7)*.

Thus, these compressive stress curves are in good agreement with models in which the ferroelectric transition is correlated with the superconducting one *(3, 7)*. However, the enhancement is relative (not absolute), which in principle also agrees with data from undoped STO in which the dielectric constant also did not diverge strongly *(29)*. Further, the relatively weak polar (ferroelectric) instability under these conditions is potentially a result of the oxygen octahedra untwisting moderately and deforming in the direction of the stress. In lightly doped STO, these effects may not be a major obstacle to enhanced superconductivity *(28)*. At higher doping, the ferroelectric modes are at higher energy (effectively hardened) than phonon modes in the lower doping state or the undoped state *(3)* and therefore the effect of enhancing superconductivity is not as dramatic as in the tensile strain case, where the oxygen octahedra are rotating in the direction that favors the ferroelectric phase instability *(30)*.

This work emphasizes that nuances are important for properly understanding the uniaxial stress and strain properties of superconducting STO. The reported anomalous changes in transition temperature cannot be explained by trivial band structure modifications or small phonon energy changes; the atomic displacements are too small to change the bands substantially or change the phonons substantially. We believe that a distinct change or a singularity in the electron and phonon properties has to take place to explain our results.

Here we reported that at strains anticipated for the ferroelectric transition, we observed a local maximum in the transition temperature (Figure 1). Moreover, by straining the material beyond the ferroelectric phase transition we detected another downturn in the critical temperature, in accordance with some theoretical expectations. These results should be also important for other low carrier concentration superconductors and also for superconductors with higher critical temperatures and structural instabilities, particularly for materials near expected structural *quantum* phase transitions and with multiple interacting orders present.

**Acknowledgements**

We thank A. Jayakody for characterizing the samples. We thank J. Cerbin for assisting with the low temperature experiments. We thank A. V. Balatsky and K. Dunnett for fruitful discussions.